\documentclass[twocolumn]{revtex4-1}

\usepackage{amsmath,amssymb}
\usepackage{graphicx}
\usepackage{dcolumn}
\usepackage{bm}
\usepackage{multirow}
\usepackage{color}

\usepackage{natbib}

\def \bsym {\boldsymbol}
\def \n {\nonumber\\}
\def \fm {\footnotemark}

\begin{document}
\title{Variationally optimized orbital approach to trions in two-dimensional materials}
\author{Yao-Wen Chang}
\email{yaowen920@gmail.com}
\author{Yia-Chung Chang}
\email{yiachang@gate.sinica.edu.tw}
\affiliation{Research Center for Applied Sciences, Academia Sinica, Taipei 11529, Taiwan}
\begin{abstract}
  In this work, trions in two-dimensional (2D) space are studied by variational method
  with trial wavefunctions being constructed by linear combinations of 2D slater-type
  orbitals (STOs). Via this method, trion energy levels and wavefunctions can be
  calculated efficiently with fairly good  accuracy. We first apply this method to study
  trion energy levels in a 2D hydrogen-like system with respect to a wide range of mass
  ratios and screening lengths. We find that the ground-state trion is bound for the whole
  parameter range, and an excited-state trion with antisymmetric permutation of electrons
  with finite angular momentum is bound for large electron-hole mass ratios or long
  screening lengths. The binding energies of ground-state trions calculated by the present
  method agree well with those calculated by more sophisticated but
  computationally-demanding methods. We then calculate trion states in various monolayer
  transition metal dichalcogenides (TMDCs) by using this method with the inclusion of
  electron-hole exchange (EHX) interaction. For TMDCs, we found that the effect of EHX can
  be significant in determining the trion binding energy and the possible existence of
  stable excited-state trions.

\end{abstract}

\maketitle

\section{Introduction}

Excitonic complexes are known to be important components in determining the optical
properties of low-dimensional semiconducting materials with direct bandgaps. A trion is a
bound state of an exciton with another charged carrier which can either be an electron or
hole. The former is called negative trion (or negatively charged exciton), and the latter
is called positive trion (or positively charged exciton). Trions were observed initially
in quantum-well systems\cite{kheng1993observation, huard2000bound}. In recent years, it
was found that trion signatures appear frequently in the optical spectroscopy of
two-dimensional (2D) materials due to their reduced dielectric screening that leads to
greatly enhanced trion binding.\cite{mak2013tightly,Lin} Such an unique situation makes 2D
materials a fertile playing ground for studying many-body physics. This subject has also
attracted a great deal of theoretical interests\cite{berkelbach2013theory,
berkelbach2017optical, durnev2018excitons}. Albeit many theoretical studies have been
done, a refined theoretical work on this topic is still desired to explain emerging
experimental observations and make better predictions.

To study the physical properties of trions in 2D materials, it is a prerequisite to know
wavefunctions and energy levels. There are three frequently-used methodologies to
calculate the trion wavefunctions in two dimensions: diagonalization of discretized trion
Hamiltonian\cite{esser2000photoluminescence, druppel2017diversity, torche2019first,
tempelaar2019many, zhumagulov2020three, zhumagulov2020trion}, quantum Monte
Carlo\cite{mayers2015binding, kylanpaa2015binding, velizhanin2015excitonic,
szyniszewski2017binding, mostaani2017diffusion}, and basis-function
expansion\cite{phelps1983ground, stebe1989ground, usukura1999stability,
ruan2000application, redlinski2001variational, hilico2002quantum, kidd2016binding,
van2017excitons, hichri2017dielectric, van2018excitons, filikhin2018trions,
kezerashvili2017trion, hichri2019charged, jan2020excited}. The first methodology includes
the diagonalization of discretized three-particle Schr\"{o}dinger equation in real
space\cite{esser2000photoluminescence} and discretized three-particle Bethe-Salpeter
equation in momentum space\cite{druppel2017diversity, torche2019first, tempelaar2019many,
zhumagulov2020three, zhumagulov2020trion}. Both methods are useful to calculate trion
energy levels but also numerically expansive to approach converged solutions. The second
methodology is considered to be accurate in solving the ground-state wavefunction of a
trion but difficult to be used in finding excited-state wavefunctions. The third
methodology has several variants based on different basis functions used. An example of
the methodology is to use a Hylleraas variational wavefunction\cite{phelps1983ground,
stebe1989ground}. The method can be vary accurate and efficient for calculating the
ground-state energy of a trion. Another basis-function approach is to use Slater-type
orbitals (STOs) to expand trion wavefunctions\cite{redlinski2001variational,
hichri2017dielectric, hichri2019charged}. The method is quite nature for quantum chemists
since a trion can be paralleled to a hydrogen ion ($H_2^{+}$ or $H^{-}$). Other examples
include 2D harmonics in complex coordinate\cite{hilico2002quantum} and hyperspherical
harmonics\cite{ruan2000application, kezerashvili2017trion, filikhin2018trions}. However,
for a more complicated potential such as the Rytova-Keldysh potential\cite{rytova,
Keldysh}, it becomes difficult to calculate two-particle integrals efficiently. One who
already overcomes the problem is the explicitly correlated Gaussian basis method, which is
commonly used in stochastic variational method\cite{usukura1999stability, kidd2016binding,
van2017excitons, van2018excitons, jan2020excited}. However, because the Gaussian basis
does not describe the exponential asymptotic behavior of wavefunctions in the current
system, the calculation requires a large number of basis functions and thus remains
computationally intensive.

While all methods mentioned above are successfully in some degrees in finding a good
ground-state trion, they may not be easily extended to the study of excited-state trions.
In particular, it is difficult to verify whether an excited-state trion with nonzero
angular momentum is bound, since the nodal structures and cusp conditions of excited-state
wavefunctions are more sophisticated than ground-state
wavefunctions\cite{kato1957eigenfunctions, scott2007nodal}. To study excited-state trions
with nonzero angular momenta, a proper set of basis functions is needed to be carefully
chosen to resemble the trion wavefunctions. STOs as basis functions are known for
applications in quantum chemistry and atomic physics\cite{zener1930analytic,
levine2009quantum}. Recently, 2D STOs are also used as basis functions to study exciton
energy levels in 2D materials\cite{wu2019exciton, henriques2020excitons,
quintela2020colloquium}. In those studies, it is shown that the exciton energy levels can
be calculated accurately by a variationally optimized procedure. Additionally, the
asymptotic behavior of STOs is known to match that of wavefunctions in few-particle
Coulombic systems in long distance and near two-particle coalescence points
\cite{kato1957eigenfunctions}. Therefore, STOs are preferred basis functions for studying
low-lying states of trions. Furthermore, these STO basis functions have a common center
that makes it easy to include the effect of Fermi-sea blocking\cite{chang2018crossover}
for electrically gated 2D materials which have become the desired experimental setup for
such studies\cite{mak2013tightly}.

In this work, we use 2D STOs as basis functions to expand exciton and trion wavefunctions.
We apply this procedure to the calculations of exciton and trion energy levels in two
dimensions. Electron-electron and electron-hole interactions in two dimensions are modeled
by Rytova-Keldysh potential\cite{rytova, Keldysh} with an adjustable screening length,
such that the present calculation can be used to simulate trion binding energies of
transition metal dichalcogenide (TMDC) monolayers and benchmark with the binding energy of
a 2D hydrogen ion. Two-particle integrals are computed by performing numerical integration
over the Fourier transform of STOs and the Rytova-Keldysh
potential\cite{chang2018crossover}. By this method, we can calculate trion energy levels
efficiently and accurately in wide ranges of electron-hole mass ratios and screening
lengths. For TMDCs, we found that the electron-hole exchange (EHX) interaction\cite{ehx}
plays an important role in determining the trion binding energy. With the the inclusion of
the EHX interaction, our calculation shows significant difference in trion binding
energies associated with dark and bright excitons in WSe$_2$, which agrees with the
experimental observation.

The article is organized as follows. In Sec.~\ref{sec:method}, we first review the exciton
and trion Hamiltonians and define the STO basis functions used. We also discuss how to
implement the EHX term for TMDCs. In Sec.~\ref{sec:numerical}, we provide the calculations
of exciton and trion energy levels with respect to different mass ratios and screening
lengths. Trion binding energies of TMDCs are also calculated and listed. A conclusion is
given in Sec.~\ref{sec:conclusion}. Details of the formulation for the variationally
optimized orbital approach are given in Appendix.

\section{Theory\label{sec:method}}

We briefly introduce exciton and trion Hamiltonians and introduce their corresponding
wavefunctions and symmetries in Sec.~\ref{sec:hamiltonian}. The variationally optimized
orbital approach to exciton and trion wavefunctions is delineated in
Sec.~\ref{sec:wavefunction}. The implementation of EHX corrections to the exciton and
trion Hamiltonians is described in Sec.~\ref{sec:EHX}.

\subsection{Exciton and trion models\label{sec:hamiltonian}}

Based on effective mass approximation, excitons can be modeled by an exciton Hamiltonian
($ \mathcal{H}_{\text{X}}$) composed of kinetic energy of an electron and a hole, and the
electron-hole interaction given by a screened Coulomb potential.
\begin{eqnarray}
  \mathcal{H}_{\text{X}}
  &=&
  \frac{|\mathbf{p}_{\text{e}}|^2}{2m_{\text{e}}}
  +\frac{|\mathbf{p}_{\text{h}}|^2}{2m_{\text{h}}}
  -{V}(|\mathbf{r}_{\text{e}}-\mathbf{r}_{\text{h}}|),
  \end{eqnarray}
where $\mathbf{r}_{e}$ ($\mathbf{r}_{h}$) denotes the coordinate of the electron (hole),
$\mathbf{p}_{i}=-\mathtt{i}\hbar \bsym{\nabla}_{i}$ is the momentum conjugate to
$\mathbf{r}_{i}$ ($i=e,h$), and $m_{\text{e}}$ ($m_{\text{h}}$) is the effective mass of
the electron (hole). $V(r)$ is the 2D screened Coulomb potential described by the
Rytova-Keldysh potential\cite{rytova, Keldysh},
\begin{eqnarray}
  V(r)
  &=&
  \frac{\pi e^2}{\kappa \rho_0}\left[H_0\left(\frac{r}{\rho_0}\right)
  -Y_0\left(\frac{r}{\rho_0}\right)\right]\n
  &=&
  \int\;e^{\mathtt{i}\mathbf{k}\cdot\mathbf{r}}
  \left[\frac{2\pi e^2}{\kappa|\mathbf{k}|\left(1+|\mathbf{k}|\rho_0\right)}\right]
  \frac{\text{d}^2k}{(2\pi)^2},
  \label{Rytova_Keldysh}
\end{eqnarray}
where $\rho_0=r_0/\kappa$, $\kappa$ is the effective dielectric constant of the host
material that includes the effect of the surrounding and $r_0$ is the screening length of
the 2D material. $H_0(r)$ and $Y_0(r)$ are the Struve function and the Bessel function of
the second kind. The second line in Eq.~(\ref{Rytova_Keldysh})gives the Fourier transform
of the potential. The Rytova-Keldysh potential reduces to the Coulomb potential $V(r)\sim
e^2/({\kappa r})$ in the long range limit ($r \gg \rho_{0}$), and in the short-range limit
($r\ll \rho_{0}$) it becomes a logarithmic potential
$V(r)\sim[e^2/(\rho_{0}\kappa)]\left[\ln({2\rho_{0}}/{r})-\gamma\right]$ with $\gamma$
being the Euler's constant\cite{mayers2015binding, quintela2020colloquium}. The
Rytova-Keldysh potential is known for its appropriate description of the dielectric
screening effect on 2D Coulomb potential and its accurate simulation of excitons and
trions in various computational studies\cite{mayers2015binding, kylanpaa2015binding,
velizhanin2015excitonic, kidd2016binding, szyniszewski2017binding, mostaani2017diffusion,
van2017excitons, zhumagulov2020three, zhumagulov2020trion, jan2020excited,
quintela2020colloquium}. Similarly, negative trions can be modeled by a trion Hamiltonian
composed of kinetic energies of two electrons and one hole, and the screened Coulomb
interaction among them
\begin{eqnarray}
\mathcal{H}_{\text{T}}
&=&
\frac{|\mathbf{p}_{\text{e}1}|^2}{2m_{\text{e}}}
+\frac{|\mathbf{p}_{\text{e}2}|^2}{2m_{\text{e}}}
+\frac{|\mathbf{p}_{\text{h}}|^2}{2m_{\text{h}}}
-{V}(|\mathbf{r}_{\text{e}1}-\mathbf{r}_{\text{h}}|)\n
&&-{V}(|\mathbf{r}_{\text{e}2}-\mathbf{r}_{\text{h}}|)
+{V}(|\mathbf{r}_{\text{e}1}-\mathbf{r}_{\text{e}2}|),
\end{eqnarray}
The Hamiltonian for positive trions can be obtained by an exchange of the  electron
effective mass and hole effective mass in the Hamiltonian.

The exciton Hamiltonian becomes separable after making a transformation in terms of the
center-of-mass (CM) coordinates $(\mathbf{R}_{\text{X}})$ and relative coordinates
$(\mathbf{r})$. We have
\begin{eqnarray}
  \mathbf{R}_{\text{X}}
  &=&
  \frac{m_{\text{e}}\mathbf{r}_{\text{e}}+m_{\text{h}}\mathbf{r}_{\text{h}}}
  {m_{\text{X}}} \hskip1ex \mbox{and} \hskip1ex
  \mathbf{r}
  =
  \mathbf{r}_{\text{e}}-\mathbf{r}_{\text{h}},
\end{eqnarray}
where $m_{\text{X}}=m_{\text{e}}+m_{\text{h}}$ is the exciton total mass. The conjugate
momenta for electron and hole can be written as
\begin{eqnarray}
  \mathbf{p}_{\text{e}}
  =
  \frac{m_{\text{e}}}{m_{\text{X}}}\mathbf{P}_{\text{X}}+\mathbf{p} \hskip1ex \mbox{and} \hskip1ex
  \mathbf{p}_{\text{h}}
  =
  \frac{m_{\text{h}}}{m_{\text{X}}}\mathbf{P}_{\text{X}}-\mathbf{p},
\end{eqnarray}
where $\mathbf{P}_X$ denotes the CM momentum and $\mathbf{p}$ is the momentum for the
electron-hole relative motion in exciton. Similarly, the trion CM and relative coordinates
are given by
\begin{eqnarray}
  \mathbf{R}_{\text{T}}
  &=&
  \frac{m_{\text{e}}\left(\mathbf{r}_{\text{e}1}+\mathbf{r}_{\text{e}2}\right)
  +m_{\text{h}}\mathbf{r}_{\text{h}}}{m_{\text{T}}},
\end{eqnarray}
\begin{eqnarray}
  \mathbf{r}_{1}=\mathbf{r}_{\text{e}1}-\mathbf{r}_{\text{h}},\hskip1ex
  \mbox{and} \hskip1ex
  \mathbf{r}_{2}=\mathbf{r}_{\text{e}2}-\mathbf{r}_{\text{h}}.
\end{eqnarray}
The coordinate transformation leads to the following relations for the conjugate momenta
\begin{eqnarray}
  \mathbf{p}_{\text{e}1}
  =
  \frac{m_{\text{e}}}{m_{\text{T}}}\mathbf{P}_{\text{T}}+\mathbf{p}_{1},\hskip1ex
  \mathbf{p}_{\text{e}2}
  =
  \frac{m_{\text{e}}}{m_{\text{T}}}\mathbf{P}_{\text{T}}+\mathbf{p}_{2},
\end{eqnarray}
\begin{eqnarray}
  \mathbf{p}_{\text{h}}
  =
  \frac{m_{\text{h}}}{m_{\text{T}}}\mathbf{P}_{\text{T}}-\mathbf{p}_{1}-\mathbf{p}_{2},
\end{eqnarray}
where $m_{\text{T}}=2m_{\text{e}}+m_{\text{h}}$ is the trion total mass. By inserting the
transformed coordinates into the Hamiltonians, we see that the CM coordinates in both the
exciton and the trion do not interact with the internal coordinates and can be neglected
in the calculations of the energy levels. With the CM coordinates being disregarded the
exciton and trion Hamiltonians can be written as
\begin{eqnarray}
  \mathcal{H}_{\text{X}}(\mathbf{r})
  &=&
  -\frac{\hbar^2\nabla^2}{2\mu_{\text{X}}}-V(r),
  \label{exciton_hamil0}
\end{eqnarray}
\begin{eqnarray}
  \mathcal{H}_{\text{T}}(\mathbf{r}_{1},\mathbf{r}_{2})
  &=&
  \mathcal{H}_{\text{X}}(\mathbf{r}_1)+\mathcal{H}_{\text{X}}(\mathbf{r}_2)
  -\frac{\hbar^2\bsym{\nabla}_{1}\cdot\bsym{\nabla}_{2}}{m_{\text{h}}}+V(r_{12}),\n
  \label{trion_hamil0}
\end{eqnarray}
where $r_{12}=|\mathbf{r}_1-\mathbf{r}_2|$ and
$\mu_{\text{X}}=(m_{\text{e}}m_{\text{h}})/(m_{\text{e}}+m_{\text{h}})$ is the exciton
reduced mass. Given the exciton and trion Hamiltonians, the exciton and trion
wavefunctions and eigenenergies can be solved from Schr\"odinger equations
$\mathcal{H}_{\text{X}}(\mathbf{r})\Psi_{\text{X}}(\mathbf{r}) =
\varepsilon_{\text{X}}\Psi_{\text{X}}(\mathbf{r})$ and
$\mathcal{H}_{\text{T}}(\mathbf{r}_{1},\mathbf{r}_{2})
\Psi_{\text{T}}(\mathbf{r}_{1},\mathbf{r}_{2}) = \varepsilon_{\text{T}}
\Psi_{\text{T}}(\mathbf{r}_{1},\mathbf{r}_{2})$, where $\Psi_{\text{X}}(\mathbf{r})$ and
$\varepsilon_{\text{X}}$ are the exciton wavefunction and eigenenergy, while
$\Psi_{\text{T}}(\mathbf{r}_{1},\mathbf{r}_{2})$ and $\varepsilon_{\text{T}}$ are the
trion wavefunction and eigenenergy.

The rotational symmetry of excitons and trions can be realized by their angular momentum
operators. The exciton and trion angular momentum operators in two dimensions are defined
as $\mathcal{L}_{\text{X}}(\mathbf{r}) \equiv
-\mathtt{i}\hbar\mathbf{e}_{\perp}\cdot\left(\mathbf{r}\times\bsym{\nabla}\right)$ and
$\mathcal{L}_{\text{T}}(\mathbf{r}_1,\mathbf{r}_2) \equiv
-\mathtt{i}\hbar\mathbf{e}_{\perp}\cdot\left(\mathbf{r}_1\times\bsym{\nabla}_1\right)
-\mathtt{i}\hbar\mathbf{e}_{\perp}\cdot\left(\mathbf{r}_2\times\bsym{\nabla}_2\right)$.
Here, $\mathbf{e}_{\perp}$ denotes the vector perpendicular to the plane of the 2D system.
It can be shown that the exciton angular momentum operator and the exciton Hamiltonian
commute. Similarly, the trion angular momentum operator and the trion Hamiltonian commute.
Therefore, the exciton and trion wavefunctions are also eigenfunctions of the exciton and
trion angular momentum operators, respectively. We write
$\mathcal{L}_{\text{X}}(\mathbf{r})\Psi_{\text{X}}(\mathbf{r}) =
l\Psi_{\text{X}}(\mathbf{r})$ and $\mathcal{L}_{\text{T}}(\mathbf{r}_1,\mathbf{r}_2)
\Psi_{\text{T}}(\mathbf{r}_{1},\mathbf{r}_{2}) =
L\Psi_{\text{T}}(\mathbf{r}_{1},\mathbf{r}_{2})$, where $l$ is the exciton angular
momentum and $L$ is the trion angular momentum.

Trion wavefunctions should also obey the rules dictated by the exchange symmetry. There
are two types of trion wavefunctions governed by the exchange rule of identical
particles\cite{sergeev2001singlet, sergeev2001triplet, courtade2017charged}
\begin{eqnarray}
  \Psi_{\text{T}}(\mathbf{r}_1,\mathbf{r}_2)
  =
  (-1)^{S}\Psi_{\text{T}}(\mathbf{r}_2,\mathbf{r}_1),
\end{eqnarray}
where $S$ is called the permutation index. For $S=0$ the trion is in a singlet state with
spatially symmetric wavefunction. For $S=1$ the trion is in a triplet state with spatially
antisymmetric wavefunction. The terms singlet and triplet are used to specify the spin
state of the pair of electrons in a negative trion, while the spin and spatial parts of
the wavefunction are factorized. Since we will not discuss total spin states of trions in
the present context, we will use symmetric and antisymmetric trions with permutation index
$S=0$ and 1 in the following discussions to avoid confusion.

\subsection{Variational wavefunctions\label{sec:wavefunction}}

While analytical exciton and trion wavefunctions are difficult to achieve, an approximate
method has to be used to numerically solve these wavefunctions. In this section, we
introduce the Rayleigh-Ritz variational method\cite{Ritz,MacDonald} to solve the problem,
in which the exciton and trion wavefunctions are expanded within a suitably chosen set of
basis functions. To fulfill the rotational symmetry for both exciton and trion, we assume
basis functions which are subjected to
$\mathcal{L}_{\text{X}}(\mathbf{r})\phi_{a}(\mathbf{r})=l_{a}\phi_{a}(\mathbf{r})$, where
$\phi_{a}(\mathbf{r})$ is a basis function and $l_a$ is the angular momentum of the basis
function. In this study, we use 2D STOs as the basis functions. A 2D STO can be written
as\cite{wu2019exciton, henriques2020excitons}
\begin{eqnarray}
  \phi_{a}(\mathbf{r})
  &=&
  \frac{e^{\mathtt{i}l_a\varphi}}{\sqrt{2\pi}}r^{n_a-1}e^{-\zeta_a r},
  \label{sto}
\end{eqnarray}
where $n_a$, $l_a$ are the principle quantum number and angular-momentum quantum number of
the orbital $\phi_a$, $\zeta_a$ is the shielding constant, and $\varphi$ is the azimuth
angle. The principle quantum number is restricted to be positive integer number and the
angular momentum is restricted to be integer number within the range $|l_a|\le l_{max}$.
For a given set of $n_a$ and $l_a$, a number of different values of $\zeta_a$ can be used
to find the optimum shape of the radial part of the wavefunction. For the calculation of
the matrix elements of mutual Coulomb interaction, it is computationally more efficient to
use the Fourier transform of 2D STOs. The Fourier transform of a 2D STO can be written as
\begin{eqnarray}
  \tilde{\phi}_{a}(\mathbf{k})
  &=&
  \int\phi_{a}(\mathbf{r})e^{-\mathtt{i}\mathbf{k}\cdot\mathbf{r}}\text{d}^2r
  =
  \frac{e^{\mathtt{i}l_a\varphi_{\mathbf{k}}}}{\sqrt{2\pi}}
  \tilde{\mathcal{R}}_{n_a,l_a}(\zeta_a,k),
\end{eqnarray}
where the radial function in momentum space can be obtained by the generating formula
\cite{chang2018crossover}
\begin{eqnarray}
  \tilde{\mathcal{R}}_{n,l}(\zeta,k)
  &=&
  \frac{2\pi(-\mathtt{i})^{n}}{k^{n+1}}\left[\frac{\text{d}^n}{\text{d}z^n}
  \frac{\left(z-\mathtt{i}\eta\sqrt{1-z^2}\right)^{|l|}}{\sqrt{1-z^2}}
  \right]_{z=\mathtt{i}{\zeta}/{k}}.\n
  \label{Rk_formula}
\end{eqnarray}
with $\eta=l/|l|$ being the sign of $l$. By using Eq.~(\ref{Rk_formula}), the radial
functions of STOs can be generated systematically with symbolic computation programs.

The variational exciton wavefunction can be written as the linear combination of the STOs
\begin{eqnarray}
  \Psi_{\text{X},i}(\mathbf{r})
  =
  \sum_{a}u_{a,i}\phi_{a}(\mathbf{r}),
  \label{exciton}
\end{eqnarray}
where $u_{a,i}$ is the exciton wavefunction coefficient. An exciton $i=(n,l)$ can be
indicated by a principle quantum number $n$ and an angular momentum $l$, with $l=l_a$
being a constant for every orbital in the exciton wavefunction. Therefore, the rotational
symmetry of exciton is preserved. The variational trion wavefunction is written as the
linear combination of products of STOs,
\begin{eqnarray}
  \Psi_{\text{T},I}(\mathbf{r}_1,\mathbf{r}_2)
  &=&
  \sum_{ab}C_{ab,I}\frac{1}{\sqrt{2}}
  \Big[\phi_{a}(\mathbf{r}_1)\phi_{b}(\mathbf{r}_2)\n
  &&+(-1)^{S}\phi_{b}(\mathbf{r}_1)\phi_{a}(\mathbf{r}_2)\Big],
 \label{trion}
\end{eqnarray}
where $C_{ab,I}$ is the trion expansion coefficient. A trion state is labeled by
$I=(N,S,L)$, including a shell number $N$, a permutation index $S$, and an angular
momentum index $L$ with $L=l_a+l_b$ being obeyed in the sum over $l_a$ and $l_b$. By using
these basis functions, the exciton and trion eigenenergies and wavefunctions can be solved
numerically. The details of derivations for matrix elements and numerical procedures are
given in Appendix~\ref{sec:appendix}.

\subsection{EHX interaction correction\label{sec:EHX}}

For many 2D materials, such as TMDCs, the EHX interaction can be a critical component to
decide the exciton and trion binding energies. Based on the many-body theory for excitons
and trions, the EHX interaction is a short-range interaction and the strength can be
derived from the matrix elements in Bethe-Salpeter equation\cite{ehx}. Additionally, the
EHX is valley-dependent, which means the strengths of EHX can be different for the
electron and hole residing at different valleys. The exciton and trion Hamiltonians
including the EHX interaction can be rewritten as
\begin{small}
\begin{eqnarray}
  \mathcal{H}_{\text{X},\tau}(\mathbf{r})
  &=&
  -\frac{\hbar^2\nabla^2}{2\mu_{\text{X}}}-V(r)+U^{x}_{\tau}\delta(\mathbf{r}),
  \label{exciton_hamil1}
\end{eqnarray}
\begin{eqnarray}
  \mathcal{H}_{\text{T}}(\mathbf{r}_{1},\mathbf{r}_{2})
  &=&
  \mathcal{H}_{\text{X},\tau_1}(\mathbf{r}_1)
  +\mathcal{H}_{\text{X},\tau_2}(\mathbf{r}_2)
  -\frac{\hbar^2\bsym{\nabla}_{1}\cdot\bsym{\nabla}_{2}}{m_{\text{h}}}
  +V(r_{12}),\n
  \label{trion_hamil1}
\end{eqnarray}
\end{small}
where the EHX interaction has been approximated by a contact potential with strength
$U^{x}_{\tau}$\cite{YCPRL}. Here the index $\tau=1$ indicates the intravalley
electron-hole pair, while $\tau=2$ indicates the intervalley electron-hole pair.  If
$\tau_1=\tau_2$, the two electrons for a negative trion are identical particles, and the
trion wavefunction is also given by Eq.~(\ref{trion}) with the quantum numbers being given
by $I=(N,S,L)$. If $\tau_1\neq\tau_2$, the two electrons are no longer identical
particles, and the trion wavefunction should be rewritten as
\begin{eqnarray}
  \Psi_{\text{T},I}(\mathbf{r}_1,\mathbf{r}_2)
  &=&
  \sum_{ab}C_{ab,I}\phi_{a}(\mathbf{r}_1)\phi_{b}(\mathbf{r}_2),
 \label{trion_1}
\end{eqnarray}
with the quantum numbers being given by $I=(N,L)$.

\section{Numerical calculations\label{sec:numerical}}

In this section, we first study the exciton and trion energy levels as functions of the
screening length ($r_0$) and mass ratio ($\sigma$). Here, we define the mass ratio as
$\sigma=m_{\text{e}}/m_{\text{h}}$. The length and energy units used here are the
effective Bohr radius, $a_0$ and effective Hartree, $\varepsilon_0$. $a_0= \kappa
a_{\text{B}}m_0/m_{\text{e}}$ with $a_{\text{B}}\simeq 0.52918$ \AA\; the Bohr radius and
$m_0$ the free electron mass. $\varepsilon_0=(2\text{Ry}/\kappa^2)m_{\text{e}}/m_{0}$ with
$\text{Ry}=13.606$ eV. Details of calculations for the matrix elements involved are given
in Appendix~\ref{sec:appendix}. By studying the dependence of energy levels on a wide
range of parameters for trions, we can examine the ability of the present method and
explore the properties of trions in extreme conditions. Then we use this method to study
excitons and trions in TMDCs and examine the effect of EHX on the trion binding energy.
The results are compared with previous theoretical results and experimental observations
reported in literature.

\subsection{Exciton energy levels}

\begin{figure}
\includegraphics[width=0.95\linewidth]{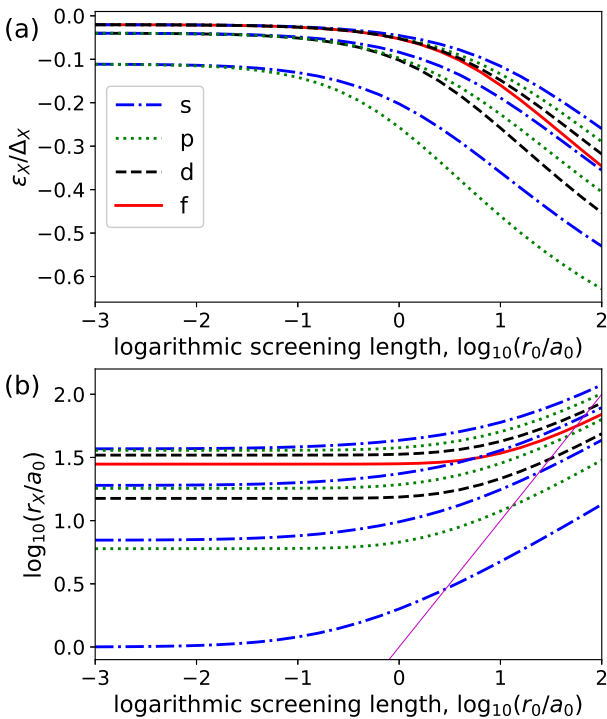}
\caption{Screening length dependence of (a) exciton energy levels and (b) exciton radii
for the case of $\sigma=1$. The exciton energies are normalized with respect to the
exciton binding energy, $\Delta_{\text{X}}$. The dash-dot lines are exciton energy levels
with $l=0$ and $n=2,3,4$ ($2s$, $3s$, $4s$ excitons) for (a) and $n=1,2,3,4$ ($1s$, $2s$,
$3s$, $4s$ excitons) for (b) from low to high. The dot lines are levels with $l=1$ and
$n=2,3,4$ ($2p$, $3p$, $4p$ excitons). The dash lines are levels with $l=2$ and $n=3,4$
($3d$, $4d$ excitons). The solid line is the level with $l=3$ and $n=4$ ($4f$ exciton).
The thin straight line denotes $r_{\text{X}}=r_0$.}
  \label{fig:exciton_dependence}
\end{figure}

Here, energy levels and average radii for the lowest few states of the exciton with
$l=0,1,2,3$ are calculated as functions of screening length and the results for mass ratio
$\sigma=1$ are shown in Fig.~\ref{fig:exciton_dependence}. All energies are normalized
with respect to the exciton binding energy $\Delta_X$. So, the exciton ground-state ($1s$)
energy is always at $-\Delta_X$ (not shown). The energy levels with $l=0,1,2,3$ ($s,p,d,
f$-like) are shown by dash-dot,  dotted, dashed, and solid curves, respectively. The
results of Fig.~\ref{fig:exciton_dependence}(a) are essentially the same as those reported
in Ref.~\onlinecite{wu2019exciton}. As $r_0\rightarrow 0$, the Rytova-Keldysh potential
reduces to the Coulomb potential in 2D, and we can benchmark the exciton energy levels
calculated by our method  with the bound-state energy levels of a 2D hydrogen atom. The
bound-state energy levels of a 2D hydrogen atom are given by the analytic
formula\cite{yang1991analytic}
\begin{eqnarray}
  \varepsilon_{\text{X},n}/\varepsilon_0=-\frac{1}{2\left(n-{1}/{2}\right)^2}\frac{1}{\sigma+1}.
\end{eqnarray}
For $\sigma=1$, the first four energy levels are $\varepsilon_{\text{H}}=-1$, $-1/9$,
$-1/25$, $-1/49$, respectively. As can be seen in Fig.~\ref{fig:exciton_dependence}(a),
the spectrum of the 2D hydrogen atom is successfully reproduced as $r_0\rightarrow{0}$. In
the other limit ($r_0/a_0\gg{1}$), the exciton spectrum becomes quite different. The
degenerate levels for each principle number are split into multiple energy levels for
different angular momenta. This is because the dynamical symmetry generated by the
Runge-Lenz vector is broken in the non-Coulombic potential for finite $r_0$. Note that in
the group of energy levels with the same principle number the level with higher absolute
value of angular momentum lies lower. For $r_0/a_0=100$, the $2p$ exciton has the binding
energy $~60$ \% of $1s$ exciton, which is in sharp contrast to the case of
$r_0/a_0\rightarrow{0}$ where the binding energy of $2p$ exciton is only about $11$ \% of
the binding energy of $1s$ exciton.

In Fig.~\ref{fig:exciton_dependence}(b), the screening length dependence of the exciton
radius is shown. The exciton radius is defined by
\begin{eqnarray}
  r_{\text{X},i}
  &\equiv&
  \frac{\int\Psi^{*}_{\text{X},i}(\mathbf{r})r\Psi_{\text{X},i}(\mathbf{r})\text{d}^2r}
  {\int\Psi^{*}_{\text{X},i}(\mathbf{r})\Psi_{\text{X},i}(\mathbf{r})\text{d}^2r}.
\end{eqnarray}
For the zero screening-length limit, it can be verified that the calculated exciton radius
is consistent with the analytical form\cite{yang1991analytic}
\begin{eqnarray}
  r_{\text{X},(n,l)}/a_0
  =
  \frac{\sigma+1}{2}\left[3n(n-1)-|l|^2+1\right]
\end{eqnarray}
with $\sigma=1$. As the screening length increases, the exciton radius also gradually
increases with a slope smaller than that for the screening length. Eventually, the
screening length exceeds exciton radius for most excitons considered. A thin straight line
for $r_{\text{X}}=r_0$ is plotted in the figure to indicate the  point of crossing for
$r_{\text{X}}$ and $r_0$. In the regime where exciton radii are shorter than the screening
length, the electron-hole attractive interaction becomes very different from the Coulomb
interaction but closer to a logarithmic interaction as mentioned in
Sec.~\ref{sec:hamiltonian}. This is consistent with the finding in the screening length
dependence of exciton energy levels in Fig.~\ref{fig:exciton_dependence} (a).

\subsection{Trion energy levels}

\begin{figure}
  \includegraphics[width=0.95\linewidth]{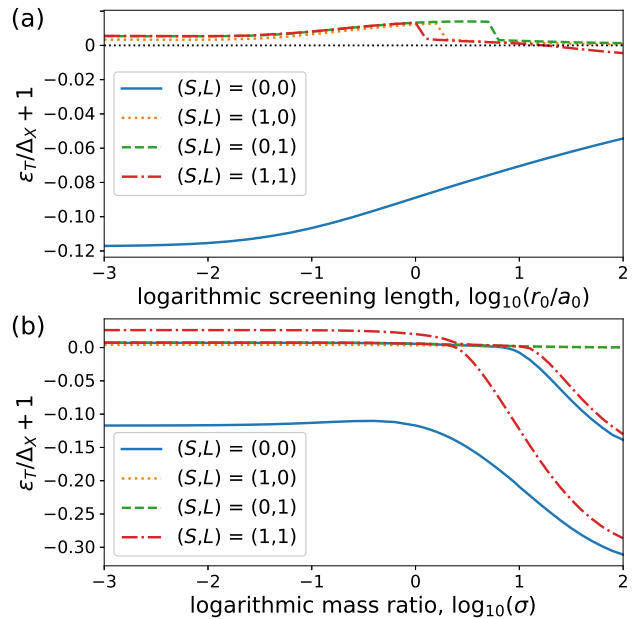}
  \caption{(a) screening length dependence of trion energy levels with $\sigma=1$ and
  $\kappa=1$; (b) mass ratio dependence of trion energy levels with $r_0=0$ and
  $\kappa=1$. The vertical axis gives the energy difference between the trion and the $1s$
  exciton in a fraction of the exciton binding energy
  ($\Delta_{\text{X}}=|\varepsilon_{\text{X},(n,l)=(1,0)}|$). For $(S,L)=(0,0)$ and
  $(S,L)=(1,1)$, the dependence of the two lowest-energy (principle number $N=1$ and
  $N=2$) trion energy levels is listed.}
  \label{fig:trion_dependence}
\end{figure}

Here, we calculate the energy levels ($\varepsilon_\text{T}$) of low-lying trion states by
varying the mass ratio ($\sigma$) and screening length ($r_0$), and the results are shown
in Fig.~\ref{fig:trion_dependence}. While only negative trions are considered explicitly,
the binding energy of positive trions can be obtained by reversing the mass ratio
($\sigma\rightarrow 1/\sigma$). It is found that, the symmetric ($S=0$) trion with zero
angular momentum ($L=0$) is bound for all the parameter regimes considered. The trion
binding energy ($\Delta_{\text{T}}=\varepsilon_{\text{X},{1s}} -\varepsilon_{\text{T}}$)
at $r_0=0$ and $\sigma=1$ is about $12.0$\% of the exciton binding energy
($\Delta_{\text{X}}=|\varepsilon_{\text{X},{1s}}|$), which is only slightly smaller than
the highest reported value $12.1$\% calculated by the stochastic variational
method\cite{usukura1999stability}. The trion binding energy at $r_0=0$ and $\sigma=0$ is
about $11.93$\% of the exciton binding energy, which is also slightly smaller than the
reported value $12.0$\%\cite{usukura1999stability}. The small discrepancy could come from
the restriction of the present variational trion wavefunction, whose asymptote behavior is
only described by a few exponential functions, while the stochastic variational method
uses hundreds of Gaussian basis functions to describe its asymptote behavior.

In addition to the trion with $(S,L)=(0,0)$, we also considered the lowest energy levels
of other sets of quantum numbers for $(S,L)$. We note that due to symmetry the trion with
angular quantum number $L=-1$ is degenerate with the trion with $L=1$ and the same
permutation index. So, we only present results for the antisymmetric ($S=1$) negative
trion with $L=1$.  It was found the negative trion with $(S,L)=(1,1)$ is bound for
$\sigma>3$ (or $\sigma<0.34$ for the positive trion) in several
literatures\cite{sergeev2001singlet, sergeev2001triplet, courtade2017charged}. As can be
seen in Fig.~\ref{fig:trion_dependence} (b), we also find the negative trion is bound for
$\sigma>2.6$ in the present calculation. For an even larger $\sigma$, there could exist
two bound states for both symmetric and antisymmetric negative trions. Additionally, as
shown in Fig.~\ref{fig:trion_dependence} (a), we find that the trion with  $(S,L)=(1,1)$
could also be bound for $\sigma=1$ and a large screening length ($r_{0}/a_{0}>20$). These
results suggest that stable antisymmetric negative trions can exist for large
electron-hole mass ratios or long screening lengths. On the other hand, for trions with
quantum numbers $(S,L)=(0,{1})$ and $(S,L)=(1,0)$, we do not find any bound state in the
present parameter range.

\subsection{Excitons and trions in TMDCs}

\begin{table*}
\centering
\caption{Trion binding energies of TMDCs}
\begin{tabular}{c | c c c c | c c c c c }
\hline
Materials & $m_{\text{e}}/m_0$ & $m_{\text{h}}/m_0$ & $r_0$(\AA) & $\kappa$
& $\Delta_{\text{X}}$(meV) & $\Delta_{\text{T}^{-}}$(meV)
& $\Delta_{\text{T}^{+}}$(meV) & $\Delta^{\star}_{\text{T}^{-}}$(meV)
& $\Delta^{\star}_{\text{T}^{+}}$(meV)\\
\hline\hline
\multirow{2}{*}{$\text{MoS}_2$}
               & \multirow{2}{*}{$0.47$} & \multirow{2}{*}{$0.54$}
               & \multirow{2}{*}{$44.68$} & $1.0$ & {\textbf{526.0}} ($526.5$\fm[1])
               & {\textbf{31.6}} ($32.0$\footnote[1]{Calculated by
               Path integral Monte Carlo method from
                Ref.~\onlinecite{kylanpaa2015binding}})
               & {\textbf{31.6}} ($31.6$\fm[1]) & \textbf{0.4} & {\textbf{2.4}} \\
               &  &  &  & $2.0$
               & {\textbf{348.4}} ($348.6$\fm[1])
               & {\textbf{24.4}} ($24.7$\fm[1])
               & {\textbf{24.5}}  { } -----  { } &-- &  -- \\
\hline
\multirow{2}{*}{$\text{MoSe}_2$}
                & \multirow{2}{*}{$0.55$} & \multirow{2}{*}{$0.59$}
                & \multirow{2}{*}{$53.16$} & $1.0$ & {\textbf{476.7}} ($476.9$\fm[1])
                & {\textbf{27.7}} ($27.7$\fm[1])
                & {\textbf{27.6}} ($27.8$\fm[1])& {\textbf{1.0}} & {\textbf{2.0}}\\
                &  &  &  & $2.0$
                & {\textbf{323.1}} ($322.9$\fm[1])
                & {\textbf{21.9}} ($22.1$\fm[1])  & {\textbf{21.9}}  { } -----  { }
                & -- & -- \\
\hline
\multirow{2}{*}{$\text{WS}_2$}
              & \multirow{2}{*}{$0.32$} & \multirow{2}{*}{$0.35$}
              & \multirow{2}{*}{$40.17$} & $1.0$ & {\textbf{508.6}} ($509.8$\fm[1])
              & {\textbf{32.4}} ($33.1$\fm[1])
              & {\textbf{32.4}} ($33.5$\fm[1])  & \textbf{0.0} & {\textbf{1.1}}\\
              &  &  &  & $2.0$
              & {\textbf{322.4}} ($322.9$\fm[1])
              & {\textbf{23.8}} ($24.3$\fm[1])
              & {\textbf{23.9}} { } -----  { } & -- & -- \\
\hline
\multirow{2}{*}{$\text{WSe}_2$}
               & \multirow{2}{*}{$0.34$} & \multirow{2}{*}{$0.36$}
               & \multirow{2}{*}{$47.57$} & $1.0$ & {\textbf{456.0}} ($456.4$\fm[1])
               & {\textbf{28.3}} ($28.5$\fm[1])
               & {\textbf{28.3}} ($28.5$\fm[1]) & \textbf{0.4} & \textbf{1.1}\\
               &  &  &  & $2.0$
               & {\textbf{294.6}} ($294.6$\fm[1])
               & {\textbf{21.3}} ($21.5$\fm[1])
               & {\textbf{21.3}} { } -----  { } & -- & -- \\
\hline
              {$\text{WSe}_2/BN (D^0)$}
              & {$0.46$} & {$0.43$}
              & {$46.8$} & $4.0$
              & {\textbf{186.0}} [$186.0$\footnote[2]{Experimentally observed exciton
               binding energy from Ref.~\onlinecite{YCPRL}}]
              & {\textbf{13.2}} [$15.2$\footnote[3]{Experimentally observed dark trion
               binding energy from Ref.~\onlinecite{liu2019gate}}]
              & {\textbf{13.2}} [$13.7$\fm[3]]& {--}& {--} \\
\hline
              {$\text{WSe}_2/BN (A^0)$}
              & {$0.38$} & {$0.43$}
              & {$46.8$} & $4.0$
              & {\textbf{172.0}} [$171.8$\footnote[2]{Experimentally observed exciton
               binding energy from Ref.~\onlinecite{liu2019magnetophotoluminescence}}]
              & {\textbf{23.0}} [$29.0$\footnote[4]{Experimentally observed bright trion
               binding energy from Ref.~\onlinecite{liu2019gate}}]
              & {\textbf{19.0}} [$21.0$\fm[4]] & -- & --\\
 \hline
\end{tabular}
\label{tab:TMDCs}
\end{table*}

We then apply the present method to calculate trion binding energies of 2D materials.
Since the EHX effect is quite significant in TMDCs, we have considered the effect of EHX
in monolayer WSe$_2$ as an example. The strengths of EHX $U^x_{\tau}$ in a contact
potential approximation for the intravalley exciton ($A^0$) and intervalley exciton
($I^0$) have been calculated via density-functional theory (DFT) in
Ref.~\onlinecite{YCPRL}. The $U^x_{\tau}$ values obtained for $A^0$ and $I^0$ are $4.17$
$\text{eV/\AA}^2$ for $\tau=1$ and $4.80$ $\text{eV/\AA}^2$ for $\tau=2$, respectively.
For comparison purpose, we first list the calculated exciton and trion binding energies
obtained by the present method with $U^x_{\tau}$ set to zero. The results for binding
energies of exciton ($\Delta_X$), negative trion ($\Delta_{T^-}$), and positive trion
($\Delta_{T^+}$) with $(S,L)=(0,{0})$ are listed in the sixth, seventh and eighth columns
in Table \ref{tab:TMDCs}, respectively. The numbers in parentheses under these three
columns are corresponding results obtained previously by using the path-integral Monte
Carlo method\cite{kylanpaa2015binding}. The binding energies of excited states of negative
trions ($\Delta_{T^-}^*$) and positive trions ($\Delta_{T^+}^*$) with $(S,L)=(1,1)$ are
also shown in the ninth and tenth columns of Table \ref{tab:TMDCs}. As shown in the Table,
the difference between the present calculation and the path-integral Monte Carlo
calculation is fairly small ($< 0.2$\%). The main resource of the discrepancy is
attributed to the small number of basis functions used in the present variational
wavefunctions, but a smaller part of the discrepancy might also come from the statistical
fluctuation of the Monte Carlo method.

We then include the EHX interaction with suitable values of $U^x_\tau$ for simulating
WSe$_2$ encapsulated by boron-nitride (BN) and calculate the exciton and trion binding
energies. The results are presented in the last row of Table \ref{tab:TMDCs}. For the case
of BN-encapsulated monolayer WSe${}_2$, accurate exciton and trion binding energies have
been measured experimentally\cite{liu2019magnetophotoluminescence, liu2019gate}, the
binding energies of trions with different valley combinations deduced from the
experimental observation are also listed (inside square brackets) in the last row of Table
\ref{tab:TMDCs}. The parameters used for the present calculations are obtained from
Ref.~\onlinecite{liu2019magnetophotoluminescence}, but slightly modified to take into
account the correction due to the EHX interaction. With the modified parameters and an
extension of basis functions to improve convergence, the energy levels of the exciton
Rydberg series for $1s$, $2s$, $3s$, and $4s$, respectively are $-172.0$, $-43.7$,
$-19.4$, and $-10.6$ meV with respect to the band gap. The results are in very good
agreement with the experiment\cite{liu2019magnetophotoluminescence}. For the positive
trion (T$^+$) associated with the bright exciton (A$^0$) we adopted $U^x_1=4.17\text{
eV/\AA}^2$\cite{YCPRL} and $U^x_2=0$, since the second hole is in $K'$ valley with
opposite spin. For both positive trion (T$^+$) and negative trion (T$^-$) associated with
the dark exciton (D$^0$) we adopted $U^x_1=0$\cite{YCPRL} and $U^x_2=4.8 \text{
eV/\AA}^2$\cite{YCPRL}, since the second electron (hole) in $K'$ valley has the same spin
as the counterpart particle in $K$ valley.  For the negative dark trion (D$^-$), the
binding energy is calculated to be $13.2$ meV, while for trion associated with bright
exciton (A$^0$), the calculated trion binding increases to $19.0$ meV for the positive
trion (A$^+$) and to $23.0 $meV for the negative trion (A$^-$). The difference is caused
by the effect of EHX interaction. For the positive dark trion (D$^+$), there is an
intervalley EHX interaction between the second hole and the electron, which leads to a
reduction in trion binding. For the positive bright trion (A$^+$), the EHX interaction
exists between the first hole and the electron, which gives a reduction in binding energy
of the bright exciton (A$^0$), while the there is no EHX interaction between the second
hole and the electron, thus the trion binding becomes enhanced. Similar scenario exists
for the negatively charged bright exciton (A$^-$) which consists of a bight exciton
(A$^0$) in the K valley and another electron in the lower conduction band in the same
valley in our consideration. As shown in Table \ref{tab:TMDCs}, there is reasonably good
agreement between calculated results and experimental observations for most cases. Note
that the energy level of the negative bright trion (A$^{-}$) observed
experimentally\cite{liu2019magnetophotoluminescence} is red shifted and split by the
Coulomb exchange between the intravalley trion and the intervalley
trion\cite{courtade2017charged}. The trion binding calculated in the present model is
significantly smaller than the observed values of $29$ meV ($35$ meV) for the $A_1^-$
($A_2^-$) peak observed in monolayer WSe$2$. To make a more meaningful comparison between
the theory and experiment, we also need to consider the intervalley exchange interaction
between the electron in the upper conduction band at the K valley and the electron in the
lower conduction band at the K' vally for the intervalley trion. Furthermore, there may be
electron traps in the sample that can lead to localized negative trions and enhance their
binding energy. These effects will be considered in future work.

Based on the present calculation, weakly-bound excited-state trions with quantum numbers
$(S,L)=(1,{1})$ for positive trions of MoS${}_2$, MoSe${}_2$, WS${}_2$, WSe${}_2$ and for
negative trions of above TMDCs except for WS${}_2$ are found. However, these excited-state
trions have not been clearly identified experimentally yet. One of the reasons could be
that the binding energies of these excited-state trions are so small that the trion
transition peaks are difficult to be distinguished from the exciton transition peaks in
optical spectra.  The present model might also be too simple to describe excited-state
trions in realistic 2D materials, since Berry-curvature effect\cite{hichri2019charged},
Coulomb exchange between charge carriers\cite{courtade2017charged}, and valley degrees of
freedom\cite{durnev2018excitons} have been neglected. It is uncertain that these
excited-state trions can still be stable in an improved model. It is a question we seek to
answer in the future.

\section{Conclusion\label{sec:conclusion}}

By using 2D STOs as basis functions and variational techniques, exciton and trion energy
levels can be solved efficiently with fairly accurate results. Ground-state and
excited-state trions are studied by the present method. The trion wavefunction and
eigenenergy are indicated by a permutation index ($S$) and an angular momentum ($L$). We
find that a weakly-bound excited-state of the  negative trion with quantum numbers
$(S,L)=(1,{1})$ could exist with a large electron-hole mass ratio ($\sigma>2.6$) or a long
screening length ($r_0/a_0>20$). The present method is also implemented to study trion
binding energies of TMDCs. The calculated binding energy of the ground-state trion is well
matched with another calculation in literature by using path-integral Monte Carlo method.
Present calculations also suggest possible existence of weakly-bound excited-state trions
with $(S,L)=(1,{1})$ for both negative and positive trions of MoS${}_2$, MoSe${}_2$,
WSe${}_2$ and the positive trion of WS${}_2$ if the dielectric screening constant is
sufficiently small. By including the EHX interaction in the calculation of exciton and
trion binding energies in monolayer WSe${}_2$ for various electronic configurations, we
find that the calculated exciton and trion binding energies agree well with the
experimentally observed values.

\section*{acknowledgment}

{This work was supported in part by the Ministry of Science and Technology (MOST), Taiwan
under Contract No. 108-2112-M-001-041 and 109-2112-M-001-046.} Y.-W.C. thanks Prof. David
D. Reichman for useful discussions and the financial support from the Postdoctoral Scholar
Program at Academia Sinica, Taiwan, ROC.

\appendix

\section{Variationally optimized orbital method\label{sec:appendix}}

\subsection{Hamiltonians in effective atomic unit\label{sec:appendix:atomic}}

The exciton and trion Hamiltonians can be written by the forms
\begin{eqnarray}
  \mathcal{H}_{\text{X}}(\mathbf{r})
  &=&
  -\frac{\sigma+1}{2}\nabla^2-V(r),
  \label{exciton_hamil2}
\end{eqnarray}
\begin{eqnarray}
  \mathcal{H}_{\text{T}}(\mathbf{r}_{1},\mathbf{r}_{2})
  &=&
  \mathcal{H}_{\text{X}}(\mathbf{r}_1)+\mathcal{H}_{\text{X}}(\mathbf{r}_2)
  -\sigma\bsym{\nabla}_{1}\cdot\bsym{\nabla}_{2}+V(r_{12}),\n
  \label{trion_hamil2}
\end{eqnarray}
where $\sigma=m_{\text{e}}/m_{\text{h}}$. The two Hamiltonians are expressed with length
unit in "$a_0$" (effective Bohr radius) and energy unit in "$\varepsilon_{0}$" (effective
Hartree), which are given by $a_0 = \kappa a_{\text{B}}{m_0}/{m_{\text{e}}}$ and
$\varepsilon_{0} = (2\text{Ry}/\kappa^2)m_{\text{e}}/{m_0}$, where $m_0$ is the free
electron mass, $a_{\text{B}}\simeq 0.52918$ \AA\; ( the Bohr radius ) and $\text{Ry}\simeq
13.606$ eV (the Rydberg). The Rytova-Keldysh potential becomes
\begin{eqnarray}
  V(r)
  &=&
  \int\;e^{\mathtt{i}\mathbf{k}\cdot\mathbf{r}}
  \left[\frac{2\pi}{|\mathbf{k}|\left(1+|\mathbf{k}|\rho_0\right)}\right]
  \frac{\text{d}^2k}{(2\pi)^2},
  \label{Rytova_Keldysh_0}
\end{eqnarray}
with $\rho_0=r_0/\kappa$.

\subsection{Matrix formulations and matrix elements}

Given the basis-function set, we still need to solve the expansion coefficients for
wavefunctions. Based on the variational principle, the expansion coefficients ($u_{a,i}$)
for exciton  states in Eq.~(\ref{exciton}) and the exciton eigenenergies
($\varepsilon_{\text{X},i}$)  can be solved from the eigenvalue equation
\begin{eqnarray}
  \sum_{b}h_{ab,\tau}u_{b,i}=\varepsilon_{\text{X},i}\sum_{b}o_{ab}u_{b,i},
  \label{exciton_eigen}
\end{eqnarray}
where $h_{ab,\tau}=t_{ab}+v_{ab}+x_{ab,\tau}$ is the the exciton Hamiltonian matrix with
$t_{ab} = -[(\sigma+1)/2]
\int\phi^*_{a}(\mathbf{r})\nabla^2\phi_{b}(\mathbf{r})\text{d}^2{r}$, the kinetic
integral, $v_{ab} = -\int\phi^*_{a}(\mathbf{r})V(r)\phi_{b}(\mathbf{r})\text{d}^2r$, the
potential integral, and $x_{ab,\tau} = [U^{x}_{\tau}/(2\pi)]\delta_{n_a,1}\delta_{ab}$ is
the exchange integral. $o_{ab} =
\int\phi^*_{a}(\mathbf{r})\phi_{b}(\mathbf{r})\text{d}^2r$ is the overlap integral.

Similarly,  the expansion coefficients ($C_{ab,I}$) for trion  states in Eq.~(\ref{trion})
and the trion eigenenergies ($\varepsilon_{\text{T},I}$) can be solved from the eigenvalue
equation
\begin{eqnarray}
  \sum_{cd}H_{ab,cd}C_{cd,I}
  =
  \varepsilon_{\text{T},I}\sum_{cd}O_{ab,cd}C_{cd,I},
  \label{trion_equation}
\end{eqnarray}
where $H_{ab,cd}$ is the trion Hamiltonian matrix and $O_{ab,cd}$ is the trion overlap
matrix. For the two electrons being identical particles, the quantum numbers of a trion is
$I=(N,S,L)$, and the trion Hamiltonian matrix and the overlap matrix are written as
\begin{eqnarray*}
  H_{ab,cd}
  &=&
  h_{ac,\tau}o_{bd}+o_{ac}h_{bd,\tau}+X_{ab,cd}+\bar{V}_{ab,cd}\n
  &&+(-1)^{S}(h_{ad,\tau}o_{bc}+o_{ad}h_{bc,\tau}+X_{ab,dc}+\bar{V}_{ab,dc}),
\end{eqnarray*}
and $O_{ab,cd} = o_{ac}o_{bd}+(-1)^{S}o_{ad}o_{bc}$, with
\begin{eqnarray*}
  X_{ab,cd}
  &=&
  -\sigma\int\phi^{*}_{a}(\mathbf{r}_1)\phi^{*}_{b}(\mathbf{r}_2)\n
  &&\times\bsym{\nabla}_{1}\cdot\bsym{\nabla}_{2}
  \left[\phi_{c}(\mathbf{r}_1)\phi_{d}(\mathbf{r}_2)\right]
  \text{d}^2r_1\text{d}^2r_2
\end{eqnarray*}
being called the "kinetic-polarization" integral and
\begin{eqnarray*}
  \bar{V}_{ab,cd}
  &=&
  \int\phi^{*}_{a}(\mathbf{r}_1)\phi^{*}_{b}(\mathbf{r}_2)V(r_{12})
  \phi_{c}(\mathbf{r}_1)\phi_{d}(\mathbf{r}_2)
  \text{d}^2r_1\text{d}^2r_2
\end{eqnarray*}
the two-particle mutual-interaction integral. On the other hand, for the case that the two
electrons reside at different valleys, the trion Hamiltonian matrix should be rewritten as
\begin{eqnarray*}
  H_{ab,cd}
  &=&
  h^{(1)}_{ac,\tau_1}o_{bd}+o_{ac}h^{(2)}_{bd,\tau_2}+X_{ab,cd}+\bar{V}_{ab,cd},
\end{eqnarray*}
and the overlap matrix becomes $O_{ab,cd} = o_{ac}o_{bd}$, while the quantum numbers are
given by $I=(N,L)$. Here  $h^{(1)}$ and  $h^{(2)}$ can be different since the effective
masses of the two electrons in the trion in TMDCs can be in different conduction bands due
to the presence of two closely-spaced conduction bands in K (K') valley.

\subsection{Orbital integrals\label{sec:integral}}

By using the STOs, analytic expressions for calculating orbital integrals can be derived.
The overlap integral is given by
\begin{eqnarray}
  o_{ab}
  &=&
  \delta_{l_a,l_b}\frac{(n_a+n_b-1)!}{(\zeta_a+\zeta_b)^{n_a+n_b}},
\end{eqnarray}
and the kinetic integral is found as
\begin{eqnarray}
  t_{ab}
  &=&
  -\frac{\delta_{l_a,l_b}(\sigma+1)}{2}
  \frac{(n_a+n_b-1)!}{\left(\zeta_a+\zeta_b\right)^{n_a+n_b}}\n
  &&\times\Bigg\{(1-\delta_{n_b,1})
  \frac{\left[(n_b-1)^2-l_b^2\right]\left(\zeta_a+\zeta_b\right)^2}
  {(n_a+n_b-1)(n_a+n_b-2)}\n
  &&-\frac{\left[(2n_b-1)\zeta_b\right]\left(\zeta_a+\zeta_b\right)}
  {(n_a+n_b-1)}+\zeta^2_b\Bigg\}.
\end{eqnarray}
By using the cross-differential operator in polar coordinates
\begin{eqnarray*}
  \bsym{\nabla}_{1}\cdot\bsym{\nabla}_{2}
  &=&
  \cos\left(\varphi_1-\varphi_2\right)
  \left(
  \frac{\partial}{\partial{r}_1}\frac{\partial}{\partial{r}_2}
  +\frac{1}{r_1}\frac{\partial}{\partial{\varphi}_1}
  \frac{1}{r_2}\frac{\partial}{\partial{\varphi}_2}\right)\n
  &&-\sin\left(\varphi_1-\varphi_2\right)
  \left(
  \frac{1}{r_1}\frac{\partial}{\partial{\varphi}_1}\frac{\partial}{\partial{r}_2}
  -\frac{1}{r_2}\frac{\partial}{\partial{\varphi}_2}\frac{\partial}{\partial{r}_1}
  \right),
\end{eqnarray*}
the kinetic-polarization integral is give by
\begin{eqnarray}
  X_{ab,cd}
  &=&
  -\frac{\sigma}{2}
  \frac{(n_a+n_c-2)!}{(\zeta_{a}+\zeta_{c})^{n_a+n_c-1}}
  \frac{(n_b+n_d-2)!}{(\zeta_{b}+\zeta_{d})^{n_b+n_d-1}}\n
  &&\times\Big[\big(\delta_{l_c,l_a-1}\delta_{l_d,l_b+1}
  +\delta_{l_c,l_a+1}\delta_{l_d,l_b-1}\big)\n
  &&\times\left(g_{ac}g_{bd}-l_cl_d\right)+\big(\delta_{l_c,l_a-1}\delta_{l_d,l_b+1}\n
  &&-\delta_{l_c,l_a+1}\delta_{l_d,l_b-1}\big)
  \left(l_dg_{ac}-l_cg_{bd}\right)\Big],
\end{eqnarray}
with $g_{ac}=n_c-1-{\zeta_{c}(n_a+n_c-1)}/{(\zeta_{a}+\zeta_{c})}$.

For the potential integral and the two-particle mutual-iteraction integral, analytical
solutions are  difficult to derive. Formulas for numerical computation are more feasible.
The potential integral can be rewritten as
\begin{eqnarray*}
  v_{ab}
  &=&
  -\int{p}_{ab}(\mathbf{r})V(r)\text{d}^2r
  =
  -\int\tilde{{p}}_{ab}(\mathbf{k})\tilde{V}(k)\frac{\text{d}^2k}{(2\pi)^2},
\end{eqnarray*}
where ${p}_{ab}(\mathbf{r})=\phi^{*}_{a}(\mathbf{r})\phi_{b}(\mathbf{r})$ is a density
matrix function with Fourier transform given by $\tilde{{p}}_{ab}(\mathbf{k})=
\int{p}_{ab}(\mathbf{r})e^{-\mathtt{i}\mathbf{k}\cdot\mathbf{r}}\text{d}^2r$ , and
$\tilde{V}({k}) = \int V(\mathbf{r})e^{-\mathtt{i}\mathbf{k}\cdot\mathbf{r}}\text{d}^2r =
{2\pi}/[k({1+ kr_*})]$ is the Fourier transform of the Rytova-Keldysh potential. Since the
density matrix function is also a STO, the Fourier transform can be solved by using
Eq.~(\ref{Rk_formula}). We have
\begin{eqnarray}
  \tilde{p}_{ab}(\mathbf{k})
  &=&
  \frac{e^{\mathtt{i}(l_b-l_a)\varphi_{\mathbf{k}}}}{2\pi}
  \tilde{\mathcal{R}}_{n_a+n_b-1,l_b-l_a}(\zeta_{a}+\zeta_{b},k),
  \label{density_matrix}
\end{eqnarray}
where $\tilde{\mathcal{R}}_{n_a+n_b-1,l_b-l_a}(2\zeta,k)$ is given  by
Eq.~(\ref{Rk_formula}). The potential integral becomes
\begin{eqnarray}
  v_{ab}
  &=&
  \frac{-\delta_{l_a,l_b}}{(2\pi)^2}\int^{\infty}_0
  \tilde{\mathcal{R}}_{n_a+n_b-1,0}(\zeta_a+\zeta_b,k)
  \tilde{V}(k)k\text{d}k.\n
  \label{potential_integral}
\end{eqnarray}
The integral in Eq.~(\ref{potential_integral}) can be computed numerically. Note that even
though the potential integral in real space can have analytical solution, the infinite
summations from the Bessel function and the Struve function still consume computational
resource. Besides, numerical integration over one-variable function is not so demanding.
The two-particle mutual-interaction integral can also be written in terms of density
matrix functions, as
\begin{eqnarray}
  \bar{V}_{ab,cd}
  &=&
  \int{p}^*_{ca}(\mathbf{r}_1)V(r_{12})
  {p}_{bd}(\mathbf{r}_2)\text{d}^2r_1\text{d}^2r_2\n
  &=&
  \frac{1}{(2\pi)^2}\int\tilde{{p}}^*_{ca}(\mathbf{k})\tilde{V}(k)
  \tilde{{p}}_{bd}(\mathbf{k})\text{d}^2k.
\end{eqnarray}
By using Eq.~(\ref{density_matrix}) and Eq.~(\ref{Rk_formula}), the above equation becomes
\begin{eqnarray}
  \bar{V}_{ab,cd}
  &=&
  \frac{\delta_{l_a-l_c,l_d-l_b}}{(2\pi)^3}\int^{\infty}_{0}
  \tilde{\mathcal{R}}^{*}_{n_a+n_c-1,\Delta{l}}(\zeta_{a}+\zeta_{c},k)\n
  &&\times\tilde{\mathcal{R}}_{n_b+n_d-1,\Delta{l}}(\zeta_{b}+\zeta_{d},k)
  \tilde{V}(k)k\text{d}k,
  \label{two_particle_integral}
\end{eqnarray}
with $\Delta{l}=l_a-l_c=l_d-l_b$. Therefore, the two-particle mutual-integral is reduced
to an one-variable integral, which can be computed efficiently via gaussian-quadrature
integration.

\subsection{Numerical procedure}

Both the exciton eigenenergy $\varepsilon_{\text{X},i}$ and the trion eigenenergy
$\varepsilon_{\text{T},I}$ are depended on the shielding constant $\zeta$ of STOs.
Therefore, by varying $\zeta$ to minimize each eigenenergy of the exciton or trion
Hamiltonian, an optimized exponential function can be found for each STO. However, if all
shielding constants and linear variational parameters are varied, the optimization problem
could be overdetermined and thus multiple local minimums would be found. It will result
slow convergence or oscillations in the optimization process. The calculation procedure
should be designed properly to avoid this problem.

Our procedure for an exciton calculation is given as follows. For the first step of an
exciton calculation, for each angular momentum $l$ we use $n_l$ STOs with $n_l$ different
principal quantum numbers to construct an exciton wavefunction and optimize one shielding
constant for each eigenstate. Subsequently, we will have $n_l$ eigenstates and $n_l$
optimized shielding constants for each eigenenergy. Then, we use these $n_l$ STOs with
different shielding constants as basis functions to construct the Hamiltonian matrix given
in in Eq.~(\ref{exciton_eigen}) and solve $n_l$ lowest eigenvalues and eigenfunctions from
the equation. The $n_l$ lowest eigenstates can be assigned as the $n_l$ lowest-lying
exciton states for the specific angular momentum $l$. Note that these $n_l$ wavefunctions
of lowest $n_l$ excitons are orthogonal mutually. To improve the binding energy, one may
add a few STOs with the same $n_l$ but different shielding constants, $\zeta_i$ by
choosing $\zeta_i$ to be the optimized shielding constant $\zeta_{n_l}$ multiplied by a
suitable scaling factor to form an even-tempered series\cite{chang2018crossover}. These
additional STOs will be needed especially for getting a convergent result for the $1s$
state when the EHX interaction is included.

The procedure for a trion calculation is similar. Each trion configuration
$(1/\sqrt{2})[\phi_{a}(\mathbf{r}_1)\phi_{b}(\mathbf{r_2})+(-1)^{S}
\phi_{a}(\mathbf{r}_2)\phi_{b}(\mathbf{r_1})]$ can be indicated by $(S,a,b)$ with
$a=(n_a,l_a,\zeta_a)$ and $b=(n_b,l_b,\zeta_b)$. The indices are restricted by $l_a+l_b=L$
with $i_a\leq{i_b}$ for $S=0$, and $i_a<i_b$ for $S=1$. For the first step of a trion
calculation, we calculate $\mathcal{N}$ eigenenergies for each state with quantum numbers
$(S,L)$ by optimizing the shielding constant $\zeta=\zeta_a=\zeta_b$ for each state. We
subsequently use all configurations with $\zeta_a$, $\zeta_b$ being given by $\mathcal{N}$
different shielding constants to construct the variational trion wavefunction. Finally, we
use the variational trion wavefunction to derive the trion eigenvalue equation as in
Eq.~(\ref{trion_equation}), and then solve $\mathcal{N}$ lowest eigenvalues and
eigenfunctions from the equation. The $\mathcal{N}$ lowest eigenvalues and eigenfunctions
can be assigned as the $\mathcal{N}$ lowest trion eigenenergies and wavefunctions.

In our calculations, STOs with principal quantum numbers ($|l|+1\leq{n}\leq 6$) and
angular momentum $|l|\leq 5$ are included to the basis set. For an exciton calculation
with the angular momentum $l$, there are $n_{l}=6-|l|$ shielding constants solved from
optimizing single-zeta exciton wavefunctions, and $n_{l}$ excitons can be found. For a
trion calculation, at least two shielding constants of the same $n_{l}$ are used to solve
the optimized trion wavefunctions. When the  EHX term is included, we added more ($~10$)
STOs basis functions for $n_l=1$ in order to get convergent result.

\end{document}